\documentclass[epj]{svjour}
\usepackage{latexsym}
\usepackage{graphics}
\usepackage{graphicx}
\usepackage{dcolumn}
\usepackage{bm}
\usepackage{url}
\usepackage{hyperref}

\usepackage{mathtools}
\usepackage{ amssymb }
\begin{document}
\title{Active Particle Condensation by Nonreciprocal and Time-delayed
Interactions}
\author{Mihir Durve\inst{1} \inst{2} \and Arnab Saha\inst{3} \and Ahmed 
Sayeed \inst{3}
}                     
\Corresponding{arnab@physics.unipune.ac.in}          
\Corresponding{sayeed@physics.unipune.ac.in}

\institute{Department of Physics, Universit\`a degli studi di Trieste, 
Trieste, Italy 34127 \and The Abdus Salam International Centre for Theoretical 
Physics, Trieste, Italy 34151 \and Department of Physics, Savitribai Phule Pune 
University, Pune, India 411007}
\date{Received: date / Revised version: date}

\abstract{ We consider flocking of self-propelling agents in two dimensions, 
each of which communicates with its neighbors within a limited vision-cone. 
Also, the communication occurs with some time-delay. The communication among 
the agents are modeled by Vicsek-rules. In this study we explore the combined 
effect of non-reciprocal interaction (induced by limited vision-cone)  among 
the agents and the presence of delay in the interactions on the dynamical 
pattern formation within the flock. We find that under these two influences, 
without any position-based attractive interactions or confining boundaries, the
agents can spontaneously condense into `drops'. Though the agents are in motion
within the drop, the drop as a whole is pinned in space. We find that this novel 
state of the flock has a well defined order and it is stabilized by the noise 
present in the system.
} 
\maketitle
\section{Introduction}
\label{intro}

The co-operative motion of {\it self-propelling} individuals or {\it {agents}}, ranging from cellular and colloidal scales
\cite{bi2016motility,zhang,darton,camley2017physical,peruani,ke,buttinoni,kudrolli,aranson,romanczuk2012active,bechinger2016active} to the scales of the 
flocks of macroscopic living entities (e.g. birds, fishes, locusts etc.)
\cite{couzin,ballerini,becco,turgut,ballerini2008interaction,buhl2006disorder,dyson2015onset,bialek2012statistical,cavagna2015flocking}  
depends on their internal degrees of freedom that consume and dissipate 
energy to their local environment in far from equilibrium
conditions. The collective dynamics of such active systems often leads to
features (e.g. dynamic pattern-formations and swarming)\cite{vicsek,chate,tu,toner,levin,gregoire,ginelli} which
are not achievable in passive systems at equilibrium, or even in 
systems driven away from equilibrium by external fields. In recent times
the research on collective dynamics of self-propelling individuals is 
contributing
significantly to our current understanding of physics of living
systems
\cite{salbreux2012actin,mayer2010anisotropies,behrndt2012forces,saha2016determining,naganathan2014active,nishikawa2017controlling}. 

In many collections of living agents on the move the main concern is to
maintain the speed and avoid collisions with other nearby agents. Both of these
requirements are achievable to an extent by adopting a simple rule, such as
`move as your neighbors do' -- that is, each agents attempts continuously to
align its heading (direction of motion) to the average heading  of its 
neighbors. Vicsek et al.
\cite{vicsek} proposed a model of velocity alignment which incorporates this
behavior. This model manifests some of the important characteristics of natural 
flocks - such as
ordered collective motion (e.g. band formation) for high agent densities and low 
noise strengths, and
uncorrelated motion for low agent densities and high noises. Some
recent reviews on this topic are \cite{zafeiris,menzel}.

Two important features of inter-agent interaction are: (1)
time delay and (2) reciprocity (or its absence). All the communications
among agents in nature must have some finite amount of delay. It is determined 
both
by the speed of the signal used to communicate and the time taken by
an agent to process the information regarding the 
communication. The communication and interaction among the agents are often
non-reciprocal in the sense that the influence of, say
agent $i$ on agent $j$ is in general different from that of the agent $j$ on
agent $i$.  A simple example is human pedestrian motion 
\cite{peacock2011pedestrian}. Each individual in a
pedestrian crowd usually notices a person in front of him, but \emph{is not 
noticed in turn by that person}.

The model we have studied here incorporates both communication delay and
non-reciprocity.  Time delay is introduced in the dynamics of the agents
by making agents `slow-to-notice' the velocities of their neighbors by one  
time step (i.e., an agent uses neighbors' velocities which are `out-of-date' by 
one time-step to determine its next step). Non-reciprocity is introduced by 
assigning a limited (i.e.  $\le 2\pi$) angular range of interaction to every 
agent. This limited angular range is termed as `vision-cone'. As an example of 
how a vision cone can induce
non-reciprocal interaction, one may consider a typical situation where, at
time $t$, the agent $j$ affects the motion of agent $i$ because $j$ is within  
the vision cone of $i$. But at the same time, it may happen that $i$ is not in 
the vision cone of $j$, and thus  $i$ does not affect the dynamics of $j$. This 
is illustrated in Fig. \ref{neighborhood}. 

In this work these two common and important features of many
natural flocks are considered together to investigate their interplay.
To our knowledge, it is by and large unexplored previously, although each of the 
features has been  
explored separately in recent studies. For example, the role of time-delay has 
been explored 
by \cite{mijalkov2016engineering} where phototactic robots are found to form 
metastable clusters depending on their inherent sensory delay (delay to sense 
the optical signal) and the intensity of the signal. Here the delay is
associated with the robots to sense an {\it{external signal}}. Time delay can
also be associated with sensing signals originated within the flock,
i.e. the {\it{internal signal}}. For example, an individual agent of a
flock can be slow enough to respond to a signal coming from its
neighbour. This causes communication-delay \cite{forgoston2008delay,hindes2016hybrid,szwaykowska2016collective,biggs2012time}. It has been shown that in
presence of a noise induced transition from translatory to rotatory
motion of agents \cite{erdmann2005noise}, the time-delayed
communication among them can introduce further instabilities where the 
harmonically interacting agents can form dynamic
clusters or swarms, with a high degree of polarity \cite{forgoston2008delay}.

Reciprocity is the norm in the interactions among passive particles (atoms,
molecules, colloidal particles etc.), though there are evidences of
non-reciprocal effective physical interactions in out-of-equilibrium
conditions. For example, when a system of particles is embedded in and interacts
with each other via a moving, out-of-equilibrium medium, the effective
interaction among the particles can be non-reciprocal. In particular,
diffusiophoretic and optical forces within colloids 
\cite{dholakia2010colloquium,shanblatt2011extended,sabass2010efficiency,soto2014self}, effective interaction among colloidal particles within a flowing 
solvent \cite{dzubiella2003depletion,khair2007motion,mejia2011bias}, shadow or
wake mediated forces between particles within flowing plasma  
\cite{tsytovich1997dust,morfill2009complex,chaudhuri2011complex} are examples 
for non-reciprocal, physical interactions. But in such cases one can recover 
the reciprocity by considering the medium and the particles together.

In contrast to many passive systems, interactions among the agents of a flock 
are often
non-reciprocal \cite{helbing2000simulating,helbing1995social}, a fact
that is barely  appreciated in the much of the literature on active
systems. The issue of non-reciprocity has been incidentally encountered (but 
not addressed specifically) in a few studies that explore the role of limited
vision-cone of active, interacting agents in a flock. Angular restriction on 
the reorientation of an individual agent within a flock alone 
can facilitate ordering\cite{gao2011angle} and can also affect the
thermodynamic character of the order-disorder transition occuring within the
flocks \cite{durve2016first}. It was also observed that 
angular restriction on interaction neighborhood in Vicsek model can reduce time 
required to reach ordered state from a disordered state\cite{tian}. Recently it 
has also been shown that vision-cone can induce
complex, self-assembled structures within the flock of self-propelling,
memory-less agents, communicating among themselves by position-based,    
attractive interactions \cite{barberis2016large}. 

In contrast to the earlier studies, here we introduce the delay in communication
among the agents and the angular restriction together, by adapting the Vicsek
model. We report here emergence of an instability leading to what we call the
drop state, which is similar in some respects to the patterns discussed
as `absorbing states of the flock with frozen 
fluctuations' in 
reports\cite{schaller2011frozen,sumino2012large,nagai2015collective}. 
In particular our study shows that the agents with narrow vision cones and 
delayed responses spontaneously condense and confine themselves within small 
regions, eventually forming a few randomly positioned, dense clusters, which we 
call `drops'.  Within such a drop, agents are in motion but the drop
as a whole is almost immobile, with only small fluctuations in the position of 
its 
center of mass. Also, the average angular momentum of the agents within the 
drop fluctuates about zero, which means these drops do not have 
votex-like dynamics either. The small size and immobility of the drop is 
sustained because individual agents perform a sequence of correlated, 
large-angle 
turn-arounds, effectively confining themselves to the drop. Importantly, we 
find that the drops are stable only if there is a finite amount of noise 
present 
in the interaction among the agents. The drop disperses if the noise strength 
falls below a critical value.  In the following sections we describe our 
model and report the results in detail.

\section{Model}
\label{sec_model}

\begin{figure} [!ht]
\includegraphics[trim=0cm 0cm 0cm 0cm, scale = 0.30]{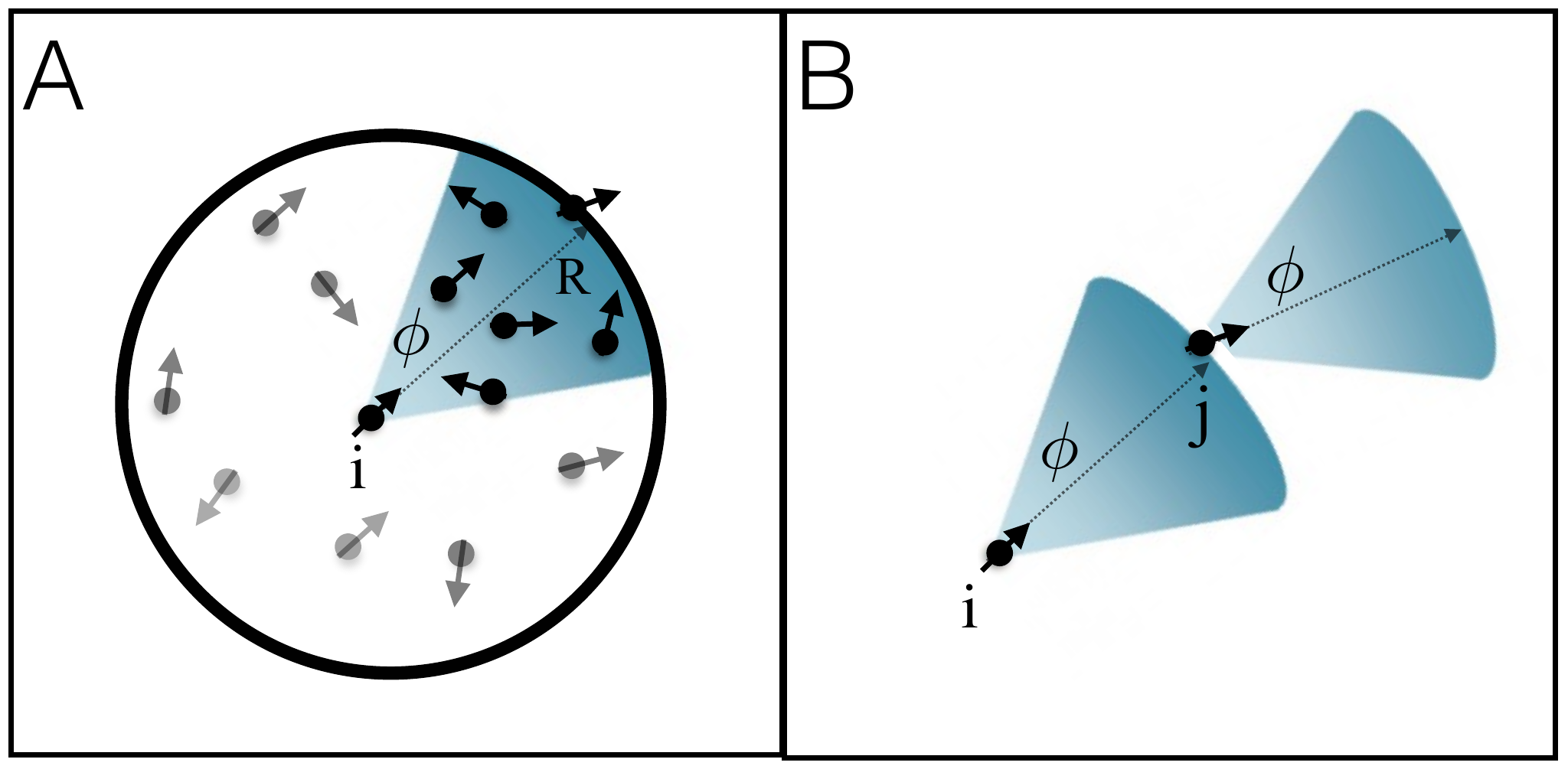}
\caption{\label{neighborhood}
(A) The neighborhood $\text{S}_i$ (blue shaded) of the $i^{\text{th}}$ agent. 
The
$i^{\text{th}}$ agent is shown at the center of a circle of radius $R$
and the neighborhood $\text{S}_i$ 
is the blue sector of the circle. The black dots with arrows as heading 
directions 
indicate the agents lying within the neighborhood (\emph{including the particle 
at the center of the circle}), and the gray 
dots with arrows as heading directions indicate agents outside it. The 
view-angle 
$\phi$ is the half opening angle of the neighborhood at the
center. (B) An example of non-reciprocal configuration of agents $i$ and
$j$, where $i$ interacts with $j$ but not the other way round.} 
\end{figure}

We add  communication-delay among the agents together with
non-reciprocity to a collection of Vicsek-agents. The non-reciprocity
is induced by the vision-cone (see the Appendix), i.e. the interaction
neighborhood of an agent which is not a circle centered around that agent (as it
was in original Vicsek model), but a sector of this circle, as illustrated 
in Fig. \ref{neighborhood}. The neighborhood sector $\text{S}_i$, which we call
the vision-cone, has an opening angle $2\phi$ and is centered about the
direction of velocity of the $i^{\text{th}}$  particle. We shall call the half
opening angle $\phi$ as the `view-angle', which can vary from $0$ to $\pi$. For
$\phi = \pi$, this model reduces to the standard Vicsek
model \cite{vicsek,chate}. 


At time $t$, $N$ number of agents have positions and velocities
$\{\mathbf{r}_i(t),\mathbf{v}_i(t)\}_{i=1}^N$.
We calculate the average velocity (denoted in square brackets below) of the
agents within
the vision-cone $S_i$ of the $i^{\text{th}}$ agent as
\begin{equation}
 [\mathbf{v}_{\text{vc}}(t)]_i =  \frac{1}{N_i} \sum \limits_{j\in{
\text{S}_i}} 
{\mathbf{v}_j}(t) 
 \label{vhatback},
\end{equation}
where $N_i$ is the number of agents within the vision-cone $S_i$ and
the subscript `vc' denotes vision-cone . Now we find
the velocity of $i^{\text {th}}$ agent at time $t+1$ as,
\begin{equation}
 \mathbf{v}_i (t+1) = v_{0} \mathcal{R}_{\eta}(\theta)\circ
[ \hat{ \mathbf{v}}_{\text{vc}}(t)]_i, 
\label{v_update}
\end{equation}
where the over-hat $\hat{}$  indicates unit vector.
$\mathcal{R}_{\eta}(\theta)$ is the
rotation operating (denoted by `$\circ$') on the unit vector
$[ \hat{ \mathbf{v}}_{\text{vc}}(t)]_i$ to rotate it by an angle $\theta$. The
angle $\theta$ is a random
variable uniformly distributed over the interval $[-\eta \pi, \eta \pi]$, where
$\eta$ is the strength (i.e., amplitude) of the noise that can be varied from $0$
to $1$. $v_0$ is the constant speed of the agents. It should be noted here that 
the neighborhood $S_i$ is as seen at time
instant $t$.

Though we calculate $\mathbf{v}_i (t+1)$, we update the positions with 
$\mathbf{v}_i (t)$, i.e.,  
\begin{equation}
 \mathbf{r}_i(t+1) =  \mathbf{r}_i(t) + \mathbf{v_{i}}(t) \Delta t,  
\label{r_update}
\end{equation}
where $\Delta t = 1$. Thus positions are updated with velocities which lag by
one time step. This introduces the delay in `response' of an agent to the
motion of its neighbors.

The degree of order in the collective motion of the particles is measured by a 
scalar order parameter $\bar\psi$ defined as, 
\begin{equation}
 \bar\psi = \left \langle\frac{1}{Nv_0} \hspace{1mm}  \hspace{1mm}\left \vert 
\sum_{i=1}^{N}
\mathbf{v}_{i}(t)\hspace{1mm} \right \vert \right\rangle,
\label{eq:orpar}
 \end{equation} 
where the angular bracket
denotes the average over multiple realizations and the steady state time
average over a time-window $\tau$. $\bar\psi$, in the perfectly
ordered state (when all the particles are moving in the same
direction) becomes unity and in the completely disordered state (when the
directions of motion are completely random) becomes zero, in the limit of $N
\rightarrow \infty$. In this report we use the phrase `ordered state' to mean
the stationary state of the system for which $\bar\psi > 0$ in the limit of $N
\rightarrow \infty$.

\section{Simulation Details}
The simulations were carried out in two-dimensional square box of size
$L$. At time $t=0$, $N$ agents are placed randomly and uniformly in the square
box. The positions of the agents are denoted by $\mathbf{r}_i ;
i=1,2 \ldots N$. The bulk agent density is given by $\rho = N/L^2$. For this 
study we carried out simulations with $N=144,256,400$ and $576$. 
Initial directions to the agents ($\theta_i ; i=1,2 \ldots N$) are assigned 
randomly and uniformly in the range $[-\pi$ to $+\pi]$.  $\theta_i$
are measured in a fixed frame of reference. All the agents have the same,
constant speed denoted by $v_0$. Then at each simulation time step the positions
$\mathbf{r}_i(t)$ and velocities  $\mathbf{v}_i(t)$ of all the agents are
updated simultaneously according to the model described in the section
\ref{sec_model}. The periodic boundary conditions are applied in both
directions.

Throughout the study the following parameters are fixed: the radius of
interaction $R=1.0$, bulk number density of agents $\rho =1.0$, and the speed
of the agents $v_0=0.5$. The steady state quantities are obtained by averaging
over $20$ independent realizations as well as over time $\tau$. Each realization 
is of
length $10^5$ time-steps, out of which we discard the initial $2\times 10^4$
time-steps and use the rest $8 \times 10^4$ time-steps for steady-state
time-averaging. Here onwards all angles are given in units of $\pi$
radians. The other specific details are provided in respective figure
captions.

\section{Results and Discussions}

\begin{figure} [!ht]
\includegraphics[trim=0cm 0cm 0cm 0cm, scale = 0.35]{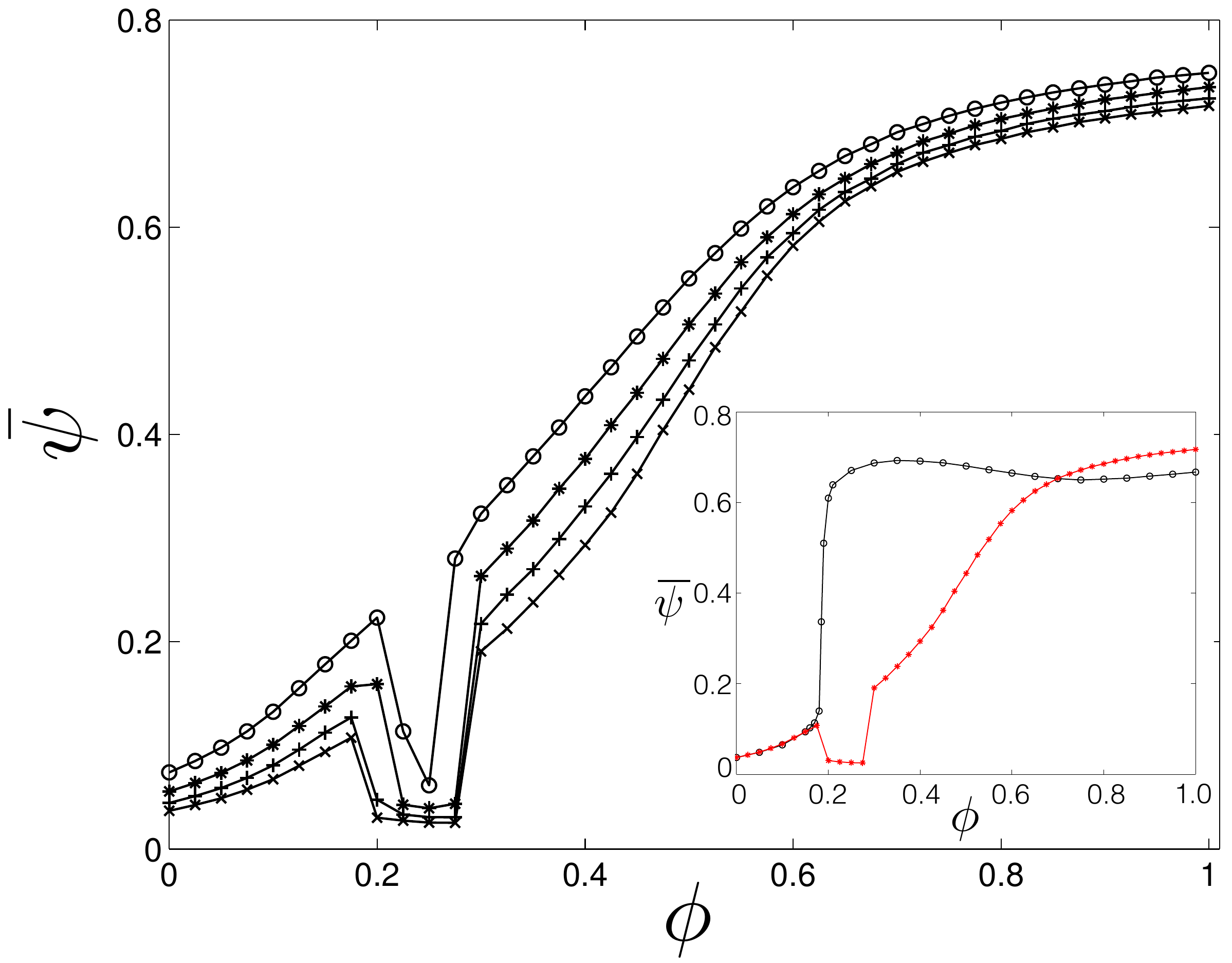}
\caption{\label{psiVsphi}
Plot of order parameter $ \bar\psi$ vs view-angle $\phi$. The circle, star,
plus and cross corresponds to system sizes $N 
= 144, 256, 400, 576$
respectively, and the noise strength $\eta = 0.3$. Inset : Plot of order parameter
$ \bar\psi$ vs view-angle $\phi$  with (red star) and without (black, open
circle) delay, N=576.} 
\end{figure} 

The main figure of Fig.[\ref{psiVsphi}] shows steady-state average (over time and
multiple configurations) of Vicsek order parameter
$ \bar\psi$ as a function of the view-angle $\phi$ for four
different system sizes. Here we see the most remarkable anomalous behavior of 
the system - the
order parameter $ \bar\psi$ dips to a value close to zero around
$\phi = 0.28$, and then again recovers to higher value at lower $\phi$ 
values. As the
system size increases, the range of $\phi$ over which the dip exists
widens. Within this anomalous range of $\phi$ ($\approx 0.20$ to
$0.28$), for $N=576$, the value of the $\bar\psi$ is slightly lower than
what is expected for a completely disordered state (which yields a small non
zero value of the order of $10^{-2}$ due to finite system size). As we shall
describe below, in this range of $\phi$, the system is indeed not in a 
disordered state,
but in a remarkable new, ordered state where the agents spontaneously confine
themselves in a small, almost immobile cluster which we will refer as a
{\emph{drop}}. Vicsek order parameter $\bar\psi$
for this drop is as close to zero as it is for completely
disordered state of the agents. Therefore $\bar\psi$ is unable to capture the 
difference
between this drop state and a completely disordered state. Later we define a new
order parameter to quantify this order.

In the inset of Fig.[\ref{psiVsphi}]  we have shown a comparison of the system
behaviour in the presence and absence of delay. In this comparison all the
parameter values in the two cases are the same, and the system without delay is
simulated using the model rules described in \cite{durve2016first}. Here we see
that in the absence of delay the drop state, which is marked by a $\bar\psi
\approx 0$ region around $\phi = 0.24$, is absent. This directly manifests the
essential role played by the delay in producing the drop state. Also in both
cases, taking the standard Vicsek dynamics limit with $\phi = 1$, we
obtain the value of order parameter
$\bar\psi \approx 0.7$. It indicates the system is in ordered state.
Therefore at this limit, in both cases (i.e., with and without delay) we
observe an ordered state with agents moving in band-like structures (not
shown here, but discussed in \cite{durve2016first}).

\begin{figure} [!ht]
\includegraphics[trim=0cm 0cm 0cm 0cm, scale = 0.32]{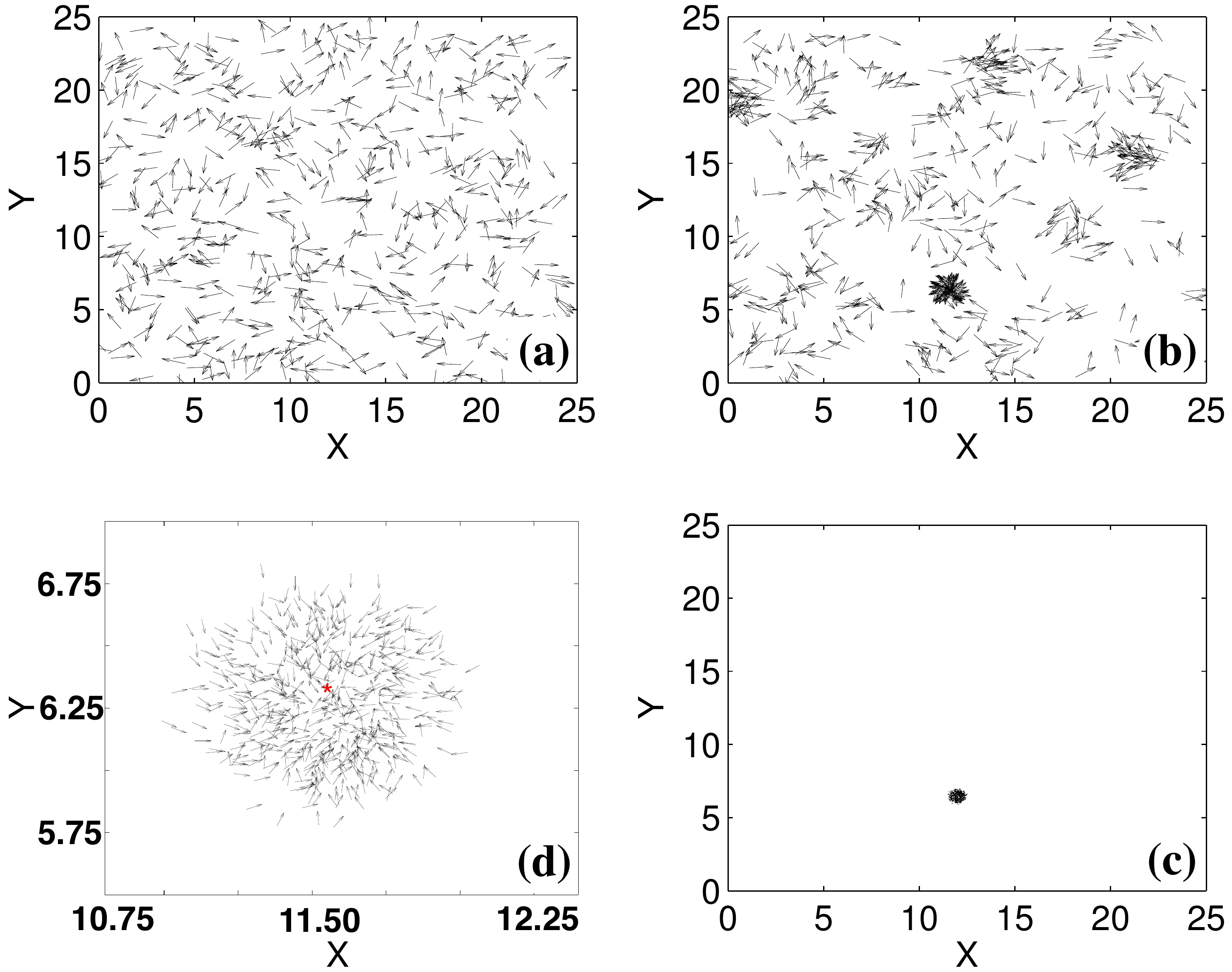}
\caption{\label{condensation}
Snapshot of the system at various time instances (a) $0$,  (b) $168$,
(c) $4500$.  Arrows indicates the directions
of the motion of the agents. (d) shows a zoomed view of the stable
drop at $t=4500$. The centre of mass of the drop is indicated by a red asterisk. 
The parameters are $N=576, \eta= 0.3$.}
\end{figure}

We now describe the emergence of this new kind of condensed state with a 
sequence of snap-shots of agent distribution within the box, shown in Fig.
\ref{condensation}. Here we set $\phi = 0.24$, because this is the mid-point of
the $\phi$-range over which the drop state is stable for the system
size $N=576$. The initial distribution in Fig.\ref{condensation}\:a shows
that the agents are homogeneously distributed 
and their directions are uniformly random. As the time progresses transient
nucleation of agents occurs at several places within the system, shown
in Fig.\ref{condensation}\:b . With time we
observe emergence of one stable nucleus shown
in Fig.\ref{condensation}\:c, which acts as a sink for the agents
passing close-by. Once a drifting agent gets within one step-length of the sink
it is irreversibly captured, and this process continues until all the agents
condense into a single drop of size roughly equal to one step-length.
This whole process can clearly be seen in the movie provided in Supplementary
Material. For the system size used for this report we obtain
only a single drop. But we have found that for larger system sizes multiple
such drops are formed spontaneously, which are stable, well-separated and 
randomly placed within the system \cite{DhurveSahaSayeed}.  
Fig. \ref{condensation}\:d  shows the zoomed view of the stable drop state of
the 
system.
The arrows indicate the
instantaneous velocity directions of the agents.  The directions appear to be
random, but there is a pronounced  radially inwards bias. This implies
that there is no
significant net tangential velocity component.  This prevents vortex-like
collective motion within the drop. 

\begin{figure} [!ht]
\includegraphics[trim=0cm 0cm 0cm 0cm, scale = 0.50]{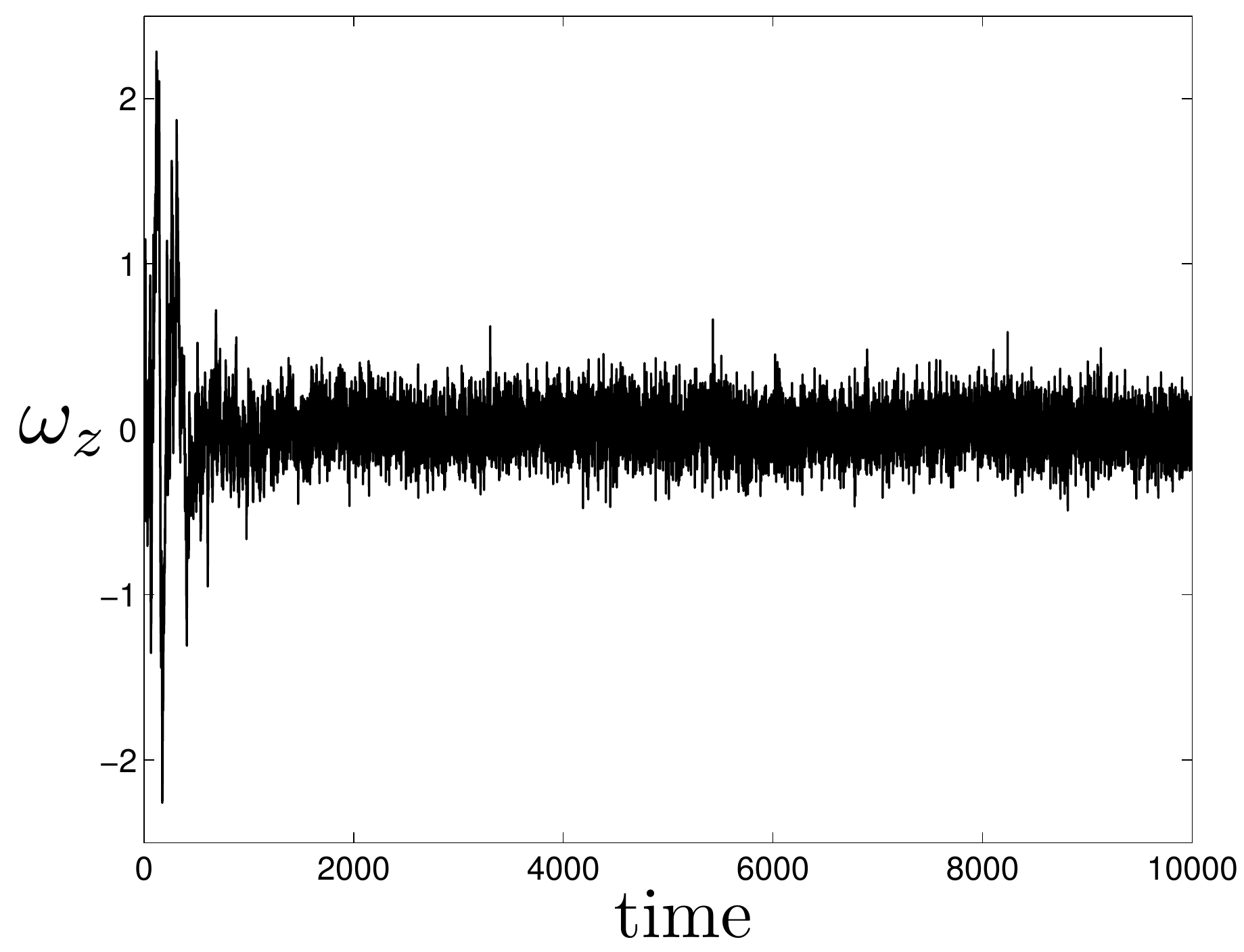}
\caption{\label{rotvel}
Average angular velocity $\omega_z$ vs $t$ plot for a single realisation. 
Parameters are $N=576, \eta=0.3,\phi=0.24$.} 
\end{figure}

The absence of vortex-like motion is also
evident from Fig.\ref{rotvel}. This figure shows the variation of average 
angular velocity
$\omega_z(t) = \langle \mathbf{r}(t) \times \mathbf{v}(t) \rangle $
with time. Here the plot is for a single realization and the
averaging is over all the agents at time $t$. Here we see that $\omega_z$ 
fluctuates rapidly about zero in time. This indicates that the agents
are equally likely to have positive or negative angular
velocity. Therefore the average angular velocity of the collection
of the agents within the drop becomes approximately zero.  

\begin{figure} [!ht]
\includegraphics[trim=0cm 0cm 0cm 0cm, scale = 0.30]{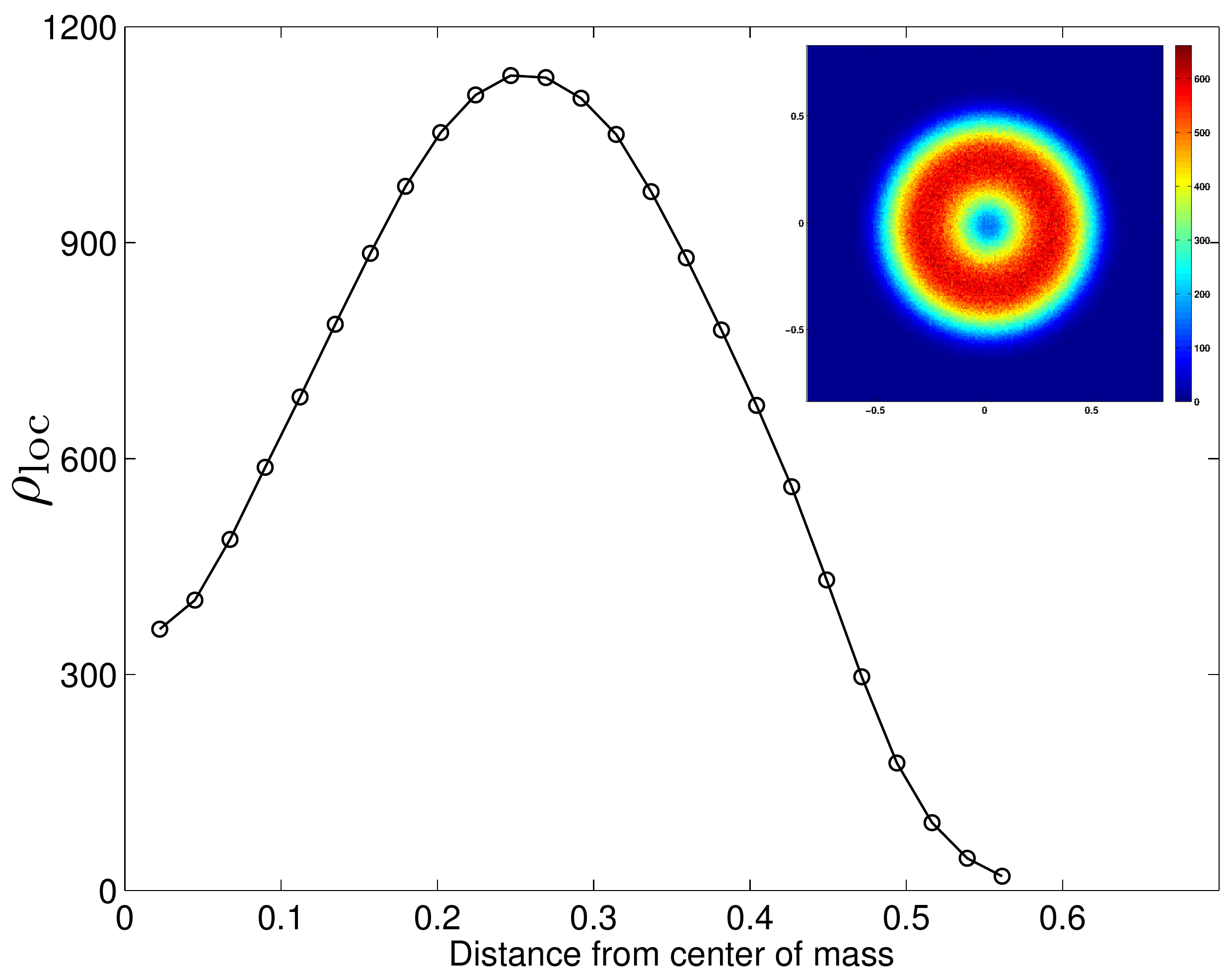}
\caption{\label{densityplot}
Radial variation of local number density of the agents from the center of mass
of the system with $N = 576$ is plotted. In the inset,
the 2d distribution of the agents within the drop of the same system is
shown in a 3d plot where the $3^{\text{rd}}$ axis (i.e. color-axis) represents
their density. The positions of the agents are measured from the
center of mass of the system, 
placed at the origin in the plot. The density plotted here is for a
single realization but averaged over time in steady state. 
The parameters are $\eta=0.3, \phi = 0.24$.} 
\end{figure}

The local number density $\rho_{\text{loc}}(r)$ (where $r$ is the
radial distance of the agents from the center of the drop) of the agents within 
the drop is
not uniform. This is the number of agents per unit area at a distance
between $r$ and $r+dr$.  It has a maximum at a certain radial distance
from the center. This is clear from Fig.\ref{densityplot}, where we see 
that the steady state density of the agents along the radius of the drop peaks
at a distance of about $0.25$, which is about half the radius of the drop.
The shape of the drop fluctuates in time, but the time-averaged shape is
circular in steady state, as is shown in the Fig.\ref{densityplot} inset. 

\begin{figure} [!ht]
\includegraphics[trim=0cm 0cm 0cm 0cm, scale = 0.65]{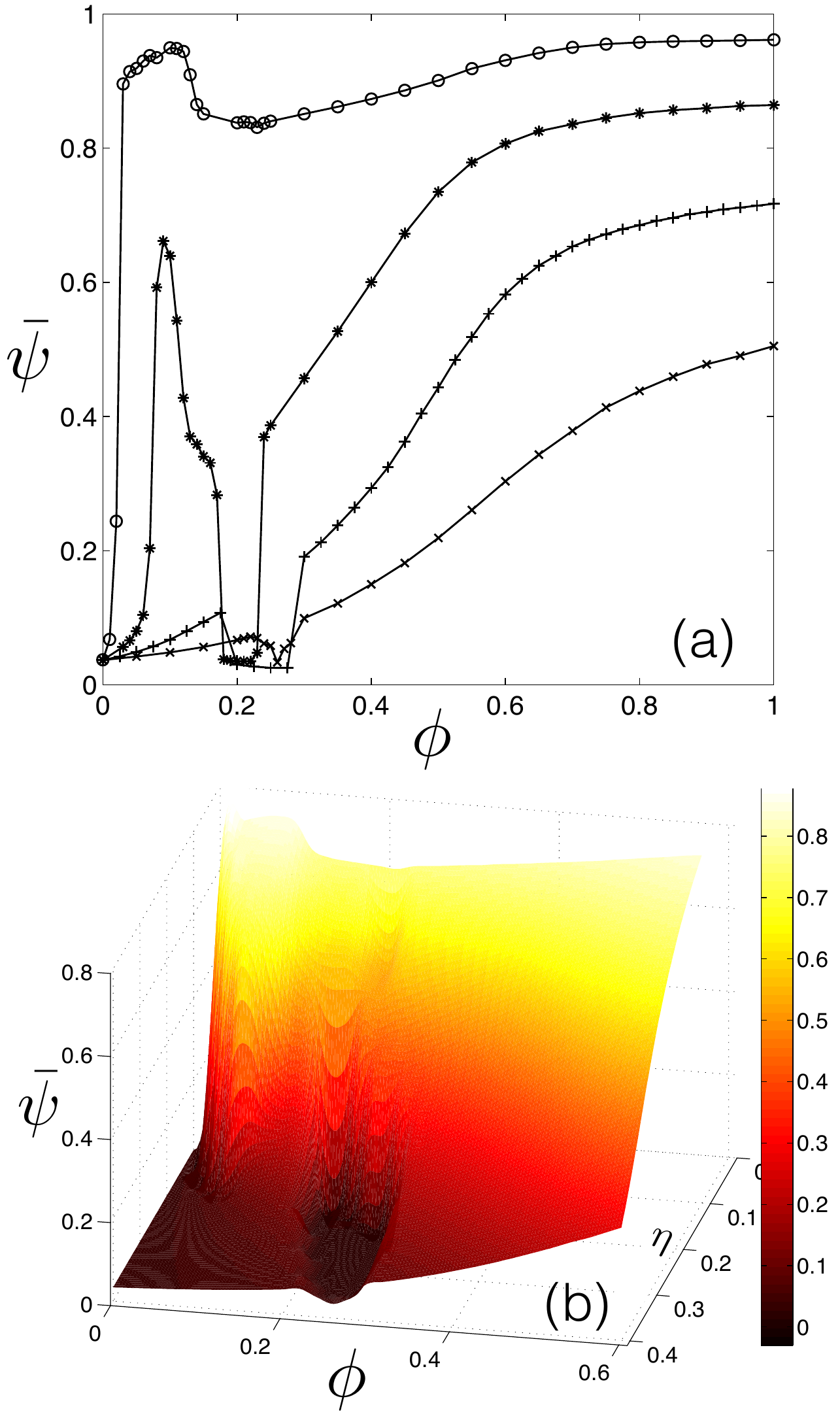}
\caption{\label{noise_order}
(a) Behavior of the order parameter $\bar\psi$ with view-angle $\phi$ for
various noise strengths. The circle, star, plus sign
and cross corresponds to noise $\eta = 0.1, 0.2, 0.3, 0.4$ respectively. Here $ N=576$. (b)Phase diagram: variation of the
order parameter $\bar\psi$ (colour axis) with view-angle $\phi$ and
noise-strength $\eta$. The valley, shown as the dark-shaded region, indicates the
parameter space for the condensation.} 
\end{figure}

\begin{figure} [!ht]
\includegraphics[trim=0cm 0cm 0cm 0cm, scale = 0.40]{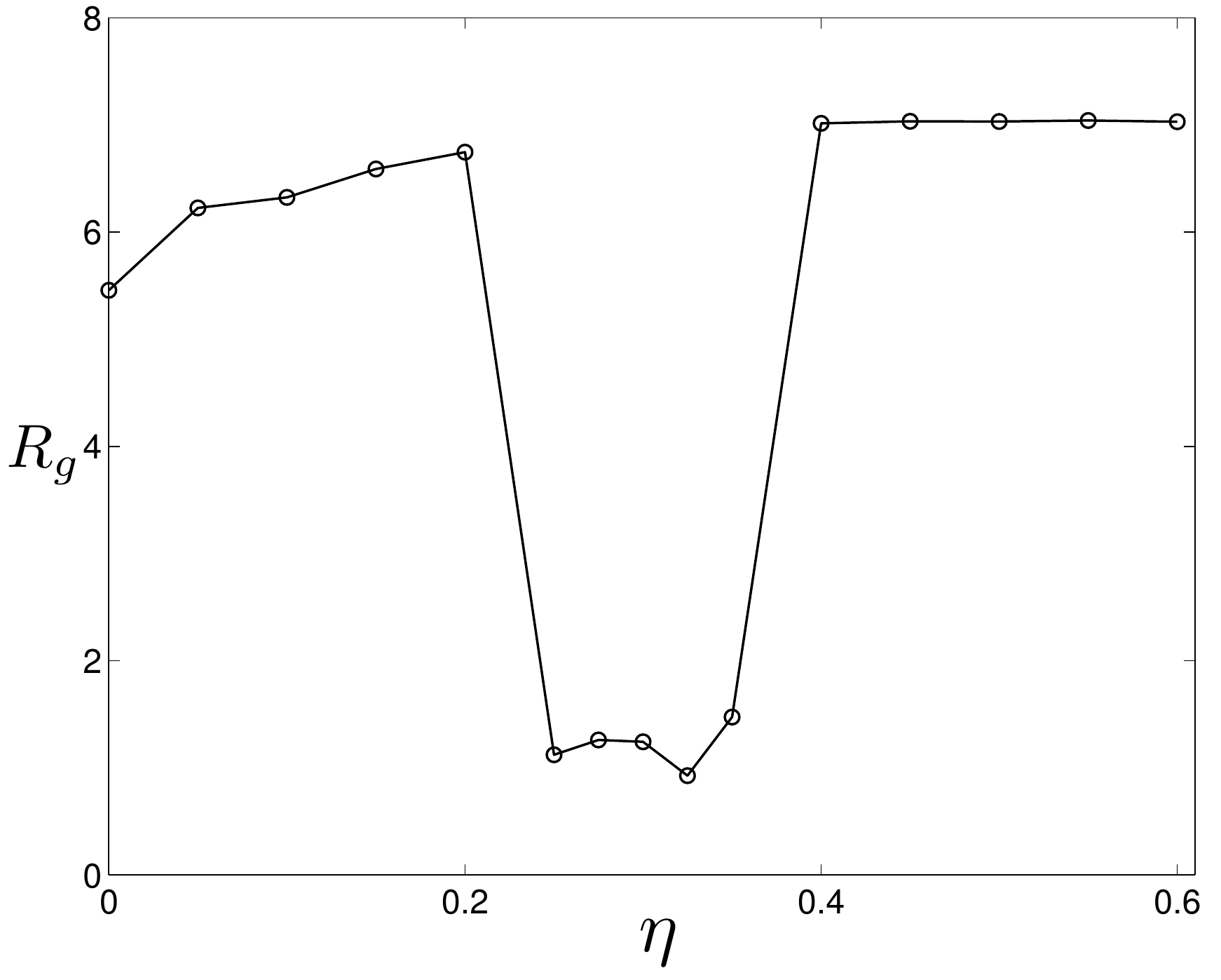}
\caption{\label{rg_order}
Radius of gyration $ Rg $ vs noise $\eta$   
plot. The parameters are : $N=576, \phi=0.24$.} 
\end{figure}

We further studied the effect of the noise on the drop. The most remarkable 
feature of the drop is that \emph{ it is stabilized by
noise}. The drop state is not stable when the noise strength is below a certain
threshold. This is shown in Fig. \ref{noise_order} and Fig. \ref{rg_order}.
Fig.\ref{noise_order}\:(a) shows the order parameter $\bar\psi$ against
view-angle $\phi$ for different noise strengths. The dip in the
order parameter (which results from the drop formation as described before) is
most pronounced for a noise $\eta = 0.3$, and is absent for $\eta \le 
0.20$ and $\eta \textgreater 0.35$. Fig.\ref{noise_order}\:(b) presents
the phase diagram over the intervals for view-angle $\phi \in (0, 0.6) $
and noise strength $\eta \in (0, 0.4)$, which manifests condensation of
agents over a considerable range of the parameters $\phi$ and $\eta$.

This condensation is also apparent in Fig. 
\ref{rg_order}, where we have shown the variation of radius of gyration $R_g$ as
a function of noise strength $\eta$.  Here $R_g$ is defined as follows:
\begin{equation}
 R_g = \left \langle \left[ \frac{1}{N} \sum_{i=1}^N (\mathbf{r}_i  -
\mathbf{r}_{ \text{cm} }  )^2  \right]^{1/2} \right \rangle . \label{eq: radgyr}
\end{equation}
In the above definition $\mathbf{r_{\text{cm}}}$ is the centre of mass of the
system and the angular brackets indicate averaging over the steady state
time-window $\tau$ and over multiple realizations. We expect $R_g$ to collapse
to a small value (comparable to the size of the drop) in the drop state, and
thus providing another signature of the transition to the drop state. In this
plot shown in Fig.\ref{rg_order} the drop-state of the system is reflected as a
small radius of gyration ($R_g \approx 1.25$), which persists over a noise range
of $\eta = 0.20 \text{--} 0.35$, and thus consistent with the observations
in Fig.\ref{noise_order}. In a different context, Chepizhko et. al. 
\cite{chepizhko2013optimal} observed that a
finite amount of noise is required to stabilize order in the
collective motion of agents in heterogeneous medium.

\begin{figure} [!ht]
\includegraphics[trim=1cm 1cm 1cm 1cm, scale = 0.40]{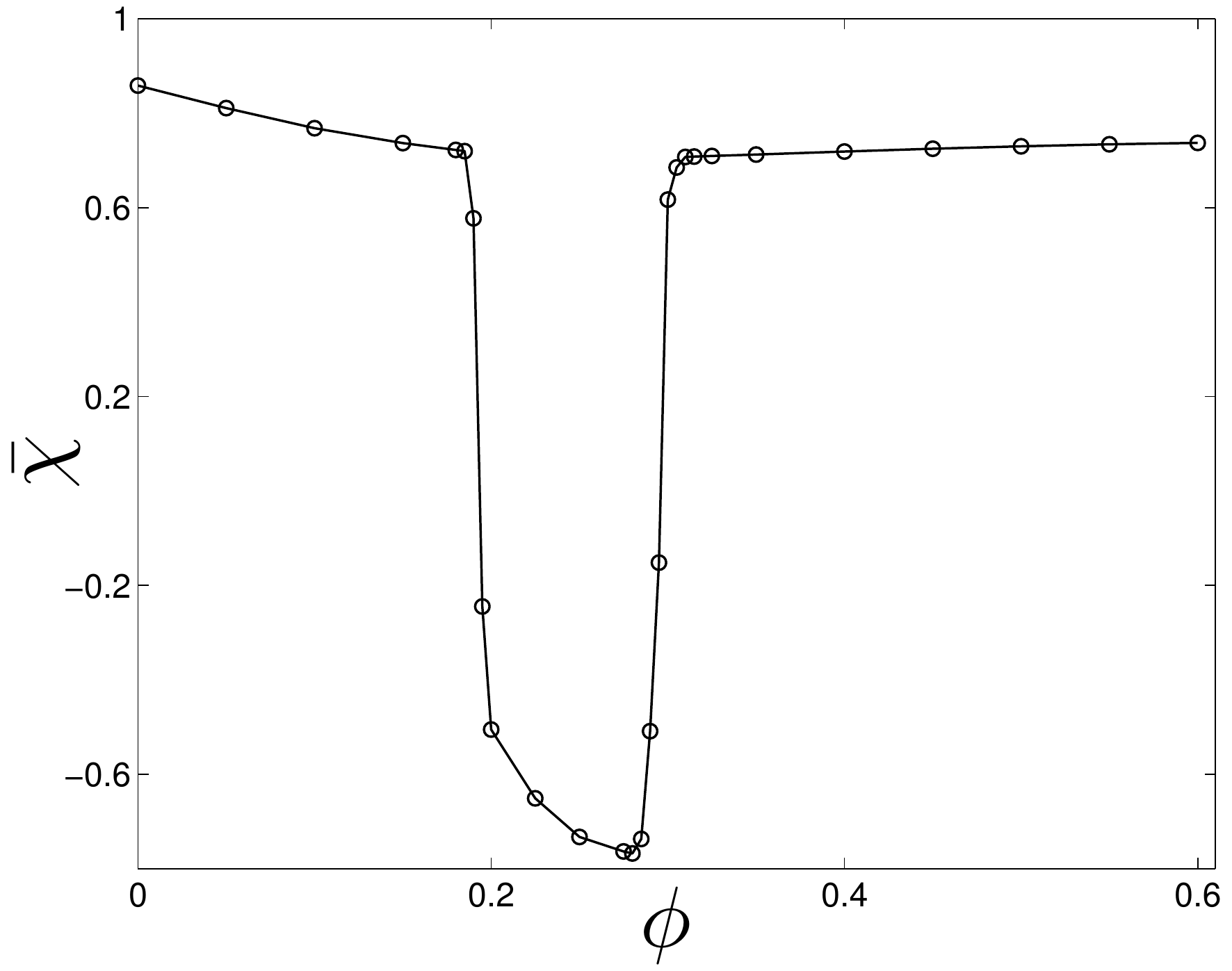}
\caption{\label{op_view_angle}
Plot of average velocity autocorrelation $ \bar\chi$ vs
view-angle $\phi$. Here $N=576$,
$\eta=0.3$.} 
\end{figure}

\begin{figure} [!ht]
\includegraphics[trim=1cm 1cm 1cm 1cm, scale = 0.45]{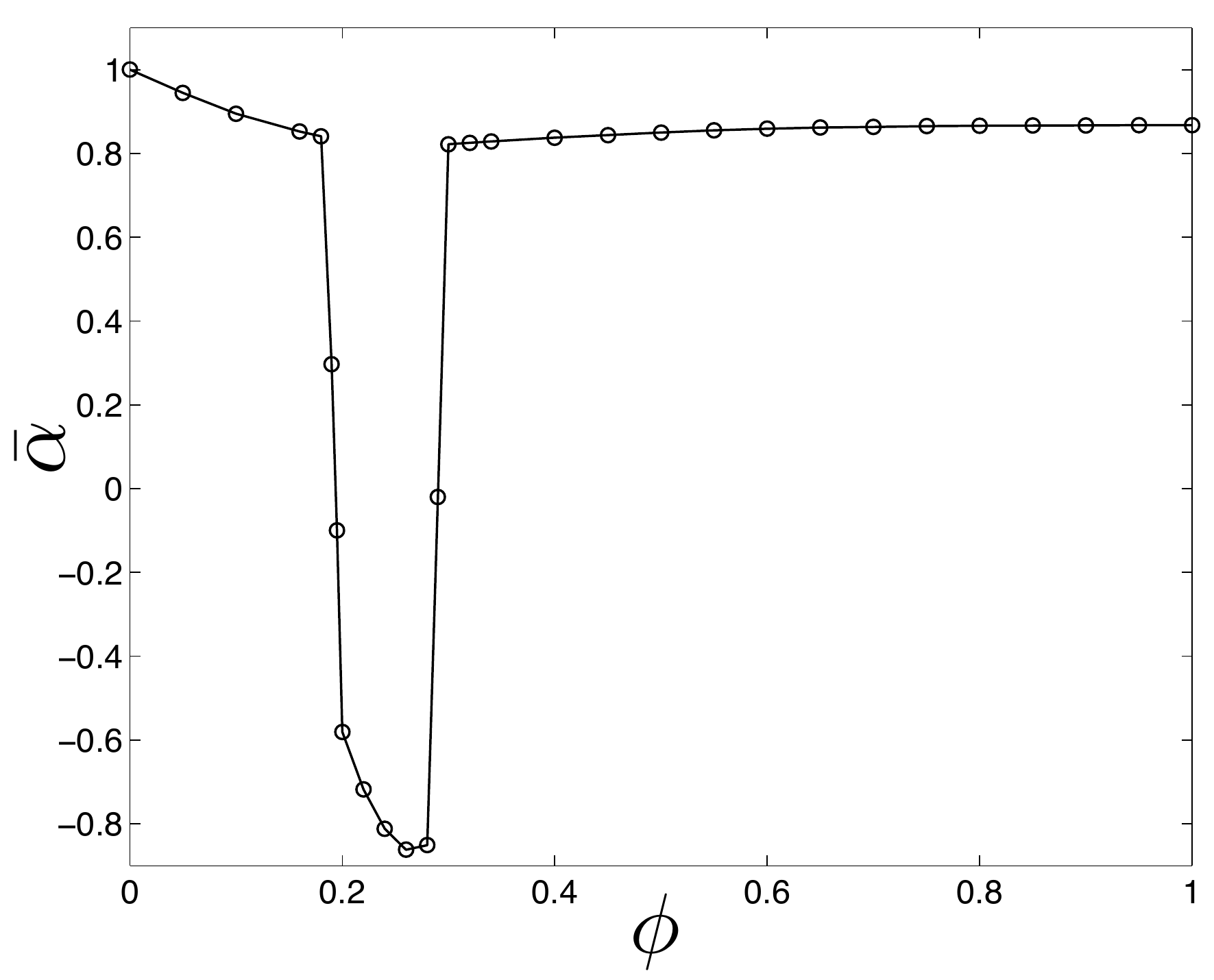}
\caption{\label{alpha} The plot of average velocity cross-correlation
$\overline{\alpha}$ versus
  view-angle $\phi$ . Here $N=576$ and $\eta=0.3\pi$.} 
\end{figure}

Now we discuss how velocity correlations of the agents vary with the
view angle $\phi$ for a given noise strength
$\eta$,  in the range where the system is in the drop state. In steady state,
we calculate average velocity \emph{autocorrelation} $\bar\chi$, which is 
defined as, 
\begin{equation}
\bar\chi = \left \langle  \frac{1}{N}\sum_i\frac{ \mathbf{v} _i(t+1) \cdot
 \mathbf{v} _i(t)}{\vert \mathbf{v} _i(t+1) \vert
\vert \mathbf{v}_i(t) 
\vert}  \right \rangle, \label{chidef}
\end{equation}
where the angular bracket implies steady state time average as well as the
average over multiple realizations. $\bar\chi$ estimates the
cosine of the angle $\Lambda$ between the directions $\hat{\mathbf{v}}_i
(t+1)$ and $\hat{\mathbf{v}}_i (t)$ of two
consecutive moves of the \emph{same} agent. We call $\Lambda$
the average turn-around angle. $\bar\chi$ is
plotted with varying view-angle in Fig.\ref{op_view_angle}. We see that
$\bar\chi\approx -0.75$ in the drop state (view-angle range: $0.18$ to
$0.3$) and otherwise $\chi \approx +0.8$.  Thus in the drop
state $\Lambda \approx 2.43$ and otherwise $\Lambda \approx
0.65$.
Also, we define a velocity cross-correlation function, between the
instantaneous velocity of the $i^{\text{th}}$ agent and the average velocity of
the agents within its vision-cone, $[\mathbf{v}_{\text{vc}}]_i$ (defined
earlier) as,
\begin{equation}
\bar \alpha =  \left\langle \frac{1}{N} \sum_i\frac{\mathbf{v}_i (t)  \cdot
[\mathbf{v}_{\text{vc}}(t)]_i
 }{|\mathbf{v}_i (t) | | [\mathbf{v}_{\text{vc}}(t)]_i | } \right \rangle .
\end{equation} 
The angular bracket indicates
the average over time in the steady
state and over multiple realizations. $\bar\alpha$ is the
average cosine of the angle $\Gamma$
between the instantaneous velocity $\mathbf{v}_i(t)$  of the $i^{\text{th}}$
agent and $ [\mathbf{v}_{\text{vc}}(t)]_i $.
Thus $\bar \alpha $ is an order parameter that measures the extent of
correlation in motion between an agent and its neighbours. As we stated earlier,
Vicsek order parameter does not capture this correlation in the motion of agents
in the drop state. But as we show here $\bar \alpha$ clearly quantifies this
correlation, as presented in Fig.\ref{alpha} which gives variation of $\bar
\alpha$ with the view-angle $\phi$.  Here we see $\bar \alpha$ has a sharp
dip within the range $0.18<\phi<0.30$, carrying the signature of the drop
formation. The value  $\bar\alpha\approx-0.85$, which corresponds to
$\Gamma \approx 2.58$. This clearly shows an agent is
moving mostly in a direction opposite to the average direction of its neighbours
in the drop state (the exact opposition in motion, i.e., $\Gamma =
\pi $ will give $\bar \alpha = - 1$). This `go against the flow' is
interesting when contrasted with usual Vicsek ordered state where agents tend to
`go along the flow'. The behaviour of $\bar \alpha$ beyond the drop phase
(i.e., outside the range $0.18<\phi<0.30$) can be explained as follows.  
For
very small view-angle $\phi$ the vision-cone of an agent contains no other
agents, so effectively the system is a collection of non-interacting agents. 
And thus $\bar \alpha$ reduces to velocity self-correlation of an agent at a 
given instant of time,
and therefore takes the value unity. For larger $\phi$ values the system attains
significant Vicsek ordering, in which case the angle between an agent's velocity
and the average velocity in the vision cone is small, i.e., $\Gamma \approx
0.63$ corresponding to $\bar \alpha \approx + 0.81$.

Taking together the observations in Fig.\ref{op_view_angle} and 
Fig. \ref{alpha} we can explain the process that spontaneously
confines the agents within the drop as follows. Each agent finds other agents in
front of it within a narrow cone and obtains their average heading (i.e. ,
$\hat{ \mathbf{v} }_{\text{vc}}$ ), which happens to be largely opposed
(i.e., by an angle $\approx 2.58 (=148^{\circ})$) to its current direction. But
the agent persists with its current heading for one time step due to the delay,
and then it turns around by a large angle (average turn-around angle 
$\Lambda\approx 2.43 (=139^{\circ})$),
remembering the average orientation of its neighbours in the previous step.
Such successive large turn-arounds effectively confine the agents to a
small region (i.e, the drop), with a linear size approximately equal to one
step-length.

\section{Finite Size Effects}

\begin{figure} [!ht]
\includegraphics[scale = 0.85]{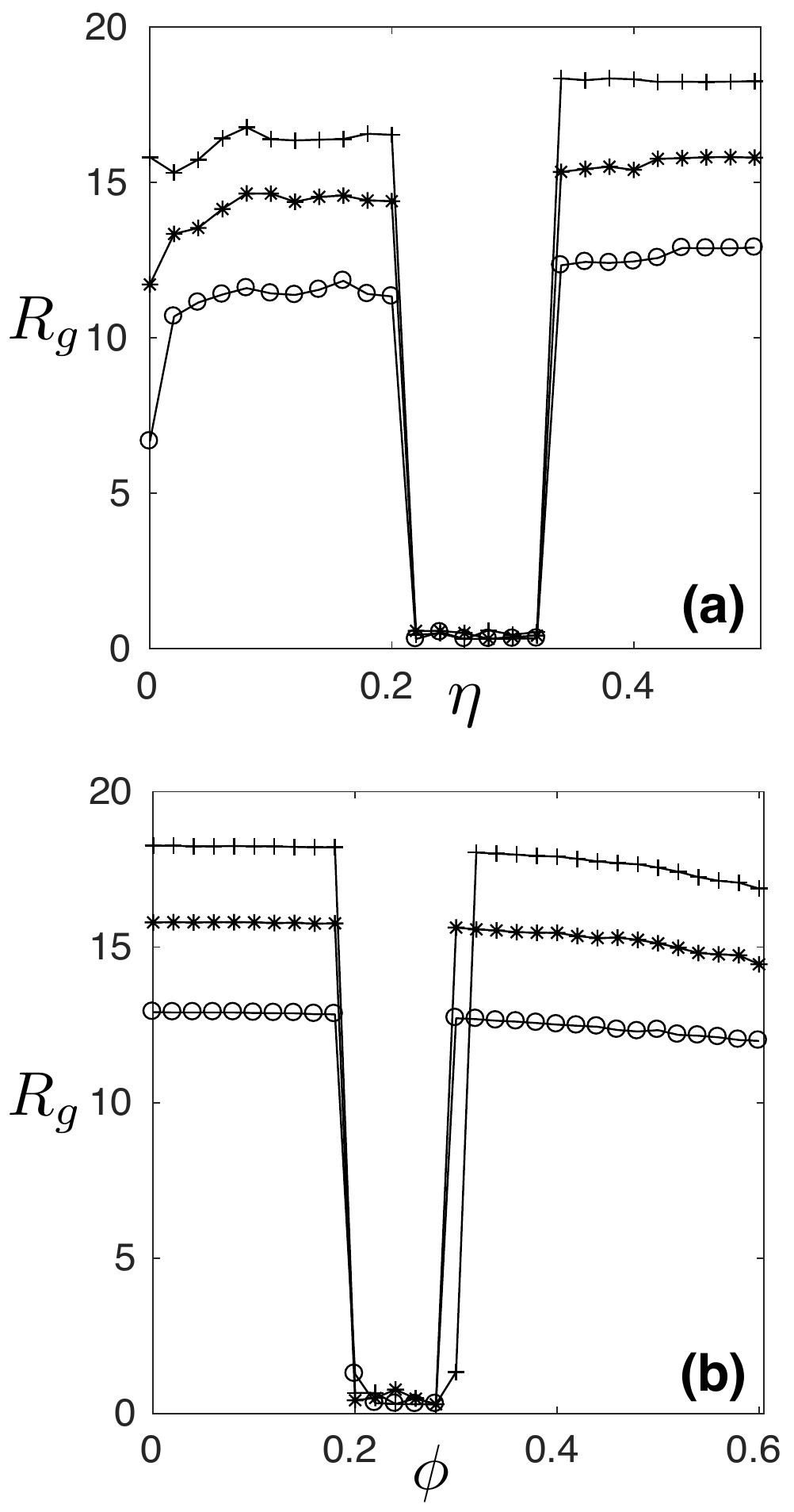}
\caption{\label{alpha1} (a) Radius of gyration $R_g$ vs Noise $\eta$. The symbol circle,star,plus corresponds to system size $N=1000,
1500, 2000$ respectively. View-angle $\phi=0.24$. (b) Radius of
gyration $R_g$ vs View-angle $\phi$. The symbol circle,star,plus
corresponds to system size $N=1000, 1500, 2000$ respectively. Noise
$\eta=0.3$.} 
\end{figure}

Here we discuss the finite size effects on the condensation, i.e. the effect of
varying the number of agents $N$ keeping the density same. We have seen that
the formation of the drop state is indicated by the collapse of $R_g$ to a
small value. Therefore we have measured the variation of $R_g$ against the noise
strength $\eta$ and the view angle $\phi$ for three system sizes $N = 1000,
1500, 2000$. This data is presented in Fig.\ref{alpha1}. Here we clearly see
that the drop state persists as we increase the system size, and the finite
size effects on the ranges of $\eta$ and $\phi$ are very small. But as we
increase the system size we find that multiple drops can form. In the
simulations for the data presented in Fig. \ref{alpha1} we found two drops in
each case. The $R_g$ values are calculated as weighted averages over the two
drops in each case.

\section{Summary and Conclusions}

We have explored the collective behavior of self-propelled
particles or agents where every agent communicates with its neighbors
through a vision cone defined by angular as well as radial limits and with a
time-delayed response to their motion. Angular limit on the vision-cone
introduces non-reciprocity in the interaction among the agents. Even in the
absence of any attractive interaction among the agents, the combination of the
non-reciprocity and the delay are enough to produce a remarkable state where
agents spontaneously condense into a dense drop. The drop remains essentially
pinned in space. It occurs within a particular parameter space of the opening
angle of the vision-cone and the noise strength of the agents, for a given bulk
agent-density of the system. In that parameter space the motion of the agents
consists of high-angle turn-arounds, which effectively confine the agents to the
drop. Within the drop the number density of the agents is radially non-uniform
and no vortex like motion is observed. Importantly, the drop  is stabilized
within a finite band of noise. With noise below than a certain strength the drop
becomes unstable. The drop is destabilized also at higher noise strength.

Now we conclude with a brief comparison between the phenomenon observed
in this study and the aggregation of active particles in the
context of motility induced phase separation (MIPS), which has been a subject of
recent research interest \cite{cates2015motility}. In both cases we observe
aggregation of active particles or
agents, but there are some crucial differences between the mechanism of the
formation of the clusters. In MIPS, a positive feed back between the
slowing-down-induced accumulation and accumulation-induced slowing
down of active particles destabilizes the uniform density state, and results in
spontaneous formation of clusters coexistent with a gas-like phase. Repulsion
or excluded-volume-interaction among the agents plays a crucial part in slowing
down mechanism within a cluster generated by MIPS. But in our case, the agents
are point-like objects, subject to Vicsek-like alignment.  Moreover, in the
aggregation we have observed (i.e, drop state), the non-reciprocal interaction
combined with time-delayed comunication among the agents are essential. To our knowledge neither of which are present in the current
theory of MIPS. Though it will be interesting to explore if one can generalize the theory of
MIPS to accommodate the non-reciprocity and delay.

\section{ACKNOWLEDGMENTS}

We acknowledge financial support from Board of College and University 
Development, Savitribai Phule Pune University. This work was carried with HPC
facilities provided by Centre for  Development of Advanced Computing (CDAC) and
also HPC facilities under the DST-FIST program  at the Department of Physics,
Savitribai Phule Pune University and HPC facilities provided by the I.C.T.P, 
Italy. One of the authors (Arnab Saha) acknowledges UGC-FRP for the financial 
support. One of the authors (Mihir Durve) is thankful for graduate fellowship 
from the I.C.T.P. Italy and University of Trieste, Italy. Authors thank Debashish Chaudhuri and
Sriram Ramaswamy for useful comments.

\section{Appendix : Vision-cone And Non-reciprocity}


Here we will discuss how a  limited vision-cone can induce non-reciprocity in
the inter-agent 
interactions. The interaction between $i^{\text{th}}$ and $j^{\text{th}}$ 
agents, namely $V_{ij}$ is non-reciprocal when $V_{ij}\neq V_{ji}$. For 
simplicity here we consider only two self-propelling agents, in zero noise 
limit, with directions of motion ${\bf {\hat v_1,\hat v_2}}$ at time $t$. 
But the discussion here can be generalised to an arbitrary number of 
agents with finite noise.
 
When the distance between the agents are less than $R$, 
the directions of motion of the agents can be 
updated at $(t+1)$ as,
 
\begin{equation}
{\bf \hat v}_1(t+1)=\frac{{\bf \hat
    v}_1(t)+\Theta(\phi-\delta_{12}){\bf \hat
    v}_2(t)}{1+\Theta(\phi-\delta_{12})}
\label{vc1}
\end{equation}

and 

\begin{equation}
{\bf \hat v}_2(t+1)=\frac{{\bf \hat
    v}_2(t)+\Theta(\phi-\delta_{21}){\bf \hat
    v}_1(t)}{1+\Theta(\phi-\delta_{21})}
\label{vc2}
\end{equation}

Here $\delta_{12}=\measuredangle{({\bf \hat v}_1,{\bf \hat
    r}_{12})}$ and $\delta_{21}=\measuredangle{({\bf \hat v}_2,{\bf \hat
    r}_{21})}$. Here $0\le(\delta_{12},\delta_{21})\le\pi$. $\Theta$ is the 
Heaviside step function
i.e. $\Theta(x)=1$ when $x\geq 0$ and $\Theta(x)=0$ if $x< 0$. If $\phi=\pi$, we 
recover Vicsek-like velocity alignment in
zero noise limit. Using the representation

\begin{equation}
 \Theta(x)=\frac{1}{2}+\frac{1}{2}\lim_{q\to\infty}\tanh(qx) \label{rep}
\end{equation}

equations \ref{vc1},\ref{vc2} can be rewritten as 

\begin{equation}
{\bf \hat v}_1(t+1)=\frac{{\bf \hat
    v}_1(t)+\frac{1}{2}(1+\lim_{q\to \infty}\tanh (q\sigma_{12})){\bf \hat
    v}_2(t)}{1+\frac{1}{2}(1+\lim_{q\to \infty}\tanh (q\sigma_{12}))}
\label{vc11}
\end{equation}   

and 

\begin{equation}
{\bf \hat v}_2(t+1)=\frac{{\bf \hat
    v}_2(t)+\frac{1}{2}(1+\lim_{q\to \infty}\tanh (q\sigma_{21})){\bf \hat
    v}_1(t)}{1+\frac{1}{2}(1+\lim_{q\to \infty}\tanh (q\sigma_{21}))}
\label{vc12}
\end{equation} 

where $\sigma_{12}=\phi-\delta_{12}$ and
$\sigma_{21}=\phi-\delta_{21}$. Here $q$ is a positive parameter signifying the
sharpness of the vision-cone boundary. Note that if agent ``2''
is within the vision-cone of the agent ``1'', $\sigma_{12}$ is
positive, and otherwise negative.  Similarly, for the agent ``1''.


For our purpose, we need the large $q$ limit (as our vision-cone boundaries are
infinitely sharp) where,\\
$\tanh(\pm q\sigma_{12}) \simeq \pm \left(1-2 \exp(-2q| \sigma_{12}|) \right)$. 
Using this asymptotic expansion when agent ``2'' is inside the vision cone of ``1'', Eq. [\ref{vc11}] becomes,


\begin{equation}
{\bf \hat v}_1(t+1)\simeq \frac{1}{2}({\bf \hat v}_1(t)+{\bf \hat
  v}_2(t))+\frac{1}{4}({\bf \hat v}_1(t)-{\bf \hat
  v}_2(t))e^{-q|\sigma_{12}|}
\label{vc21}
\end{equation}   

In Eq. \ref{vc21} The first term is Vicsek-like aligning term, and the second
term corresponds to the finite but large sharpness of vision-cone boundary. It
is straight forward to see that similar expression can be obtained for ${\bf
  \hat v}_2(t+1)$ from Eq. \ref{vc12} using the asymptotic expansion,
where $\sigma_{12}$ will be replaced by $\sigma_{21}$ as follows,

\begin{equation}
{\bf \hat v}_2(t+1)\simeq \frac{1}{2}({\bf \hat v}_2(t)+{\bf \hat
  v}_1(t))+\frac{1}{4}({\bf \hat v}_2(t)-{\bf \hat
  v}_1(t))e^{-q|\sigma_{21}|}
\label{vc22}
\end{equation} 
In general $\delta_{12}$ is independent of $\delta_{21}$, and therefore 
$\sigma_{12}\neq\sigma_{21}$. Thus, though the first term of
Eqs. \ref{vc21},\ref{vc22}, i.e. the Vicsek-like term is symmetric
under the exchange of the agents, the second term is not. And this is
the root cause for non-reciprocity. Eq. \ref{vc21} and \ref{vc22}
can be written in a concise way for any time-step $\Delta t$ as,

\begin{eqnarray}
\frac{\Delta{\bf \hat v}_i}{\Delta t}&=&\frac{({\bf \hat v}_j-{\bf \hat
    v}_i)}{2\Delta
  t}\left(1-\frac{1}{2}e^{-q|\sigma_{ij}|}\right)\\
 &=&{\bf F}_{ij}\left(1-\frac{1}{2}e^{-q|\sigma_{ij}|}\right)
\label{vc23}
\end{eqnarray} 
where $(i,j)\in(1,2)$ and $i\neq j$. For the above dynamics, ${\bf
  F}_{ij}=-{\bf F}_{ji}$ and this can be considered as the reciprocal force
acting between $i^{\text{th}}$ and $j^{\text{th}}$ agents.  The 
non-reciprocity arises from
the fact that $\sigma_{ij}\neq\sigma_{ji}$. Here we like to point out that the
non-reciprocity persists for  arbitrarily large but finite value of
the sharpenss parameter $q$.
$q|\sigma_{ij}|$
In $q|\sigma_{ij}|\to\infty$ limit we recover
the reciprocal Vicsek-like interaction. The interaction above in 
Eq.\ref{vc23} is similar to the one used to simulate a passive 
non-reciprocal system in \cite{ivlev2015statistical}.

When agent ``2'' is not inside the vision-cone of ``1'', Eq. \ref{vc11}]
becomes,  
\begin{equation}
{\bf \hat v}_1(t+1)\simeq {\bf \hat v}_1(t)+({\bf \hat v}_2(t)-{\bf
  \hat v}_1(t))e^{-q|\sigma_{12}|}
\label{vc24}
\end{equation}   
Similar expressions can also be obtained for ${\bf \hat v}_2(t+1)$
from Eq.\ref{vc12} where instead $\sigma_{12}$, we will have $\sigma_{21}$.
As $\sigma_{ij}\neq\sigma_{ji}$ in general, in this case also the
system is in general non-reciprocal. In the limit of
$(q|\sigma_{ij}|,q|\sigma_{ji}|)\to \infty$ the system becomes
non-interacting and therefore trivially reciprocal.


\bibliographystyle{ieeetr}
\bibliography{frozen}

\end{document}